\documentclass[10pt,onecolumn]{IEEEtran}
\usepackage[latin1]{inputenc}
\usepackage{amsmath}
\usepackage{amsbsy}
\usepackage{amssymb}
\usepackage{latexsym}
\usepackage{times,color}
\usepackage{epsfig}
\title{Sparsity-Aware Adaptive Algorithms Based on Alternating Optimization with Shrinkage }

\author{Rodrigo C. de Lamare and Raimundo Sampaio-Neto  \vspace{-0.25em} \\
\thanks{Copyright (c) 2012 IEEE. Personal use of this material is permitted.
Prof. R. C. de Lamare is with CETUC-PUC-Rio, 22453-900, Rio de Janeiro,
Brazil, and with the Communications Research Group, Department of
Electronics, University of York, York Y010 5DD, United Kingdom and
Prof. R. Sampaio-Neto is with CETUC/PUC-RIO, 22453-900, Rio de
Janeiro, Brazil. E-mails: rcdl500@york.ac.uk,
raimundo@cetuc.puc-rio.br} }

\begin{document}

\maketitle

\begin{abstract}
This letter proposes a novel sparsity-aware adaptive filtering
scheme and algorithms based on an alternating optimization strategy
with shrinkage. The proposed scheme employs a two-stage structure
that consists of an alternating optimization of a
diagonally-structured matrix that speeds up the convergence and an
adaptive filter with a shrinkage function that forces the
coefficients with small magnitudes to zero. We devise alternating
optimization least-mean square (LMS) algorithms for the proposed
scheme and analyze its mean-square error. Simulations for a system
identification application show that the proposed scheme and
algorithms outperform in convergence and tracking existing
sparsity-aware algorithms.

\end{abstract}

\section{Introduction}

In the last few years, there has been a growing interest in adaptive
algorithms that can exploit the sparsity present in various signals
and systems that arise in applications of adaptive signal processing
\cite{gu}-\cite{jidf}. The basic idea is to exploit prior knowledge
about the sparsity present in the data that need to be processed for
applications in system identification, communications and array
signal processing. Several algorithms based on the least-mean square
(LMS) \cite{gu,chen} and the recursive least-squares (RLS)
\cite{babadi,angelosante,eksioglu,eksioglu2} techniques have been
reported in the literature along with different penalty or shrinkage
functions. These penalty functions perform a regularization that
attracts to zero the coefficients of the adaptive filter that are
not associated with the weights of interest. With this objective in
mind, several penalty functions that account for the sparisty of
data signal have been considered, namely: {an approximation of the
$l_0$-norm \cite{gu,eksioglu2}, the $l_1$- norm penalty
\cite{chen,eksioglu}, and the log-sum penalty
\cite{chen,eksioglu,candes}}. These algorithms solve problems with
sparse features without relying on the computationally complex
oracle algorithm, which requires an exhaustive search for the
location of the non-zero coefficients of the system. However, the
available algorithms in the literature also exhibit a performance
degradation as compared to the oracle algorithm, which might affect
the performance of some applications of adaptive algorithms.

Motivated by the limitation of existing sparse adaptive techniques,
we propose a novel sparsity-aware adaptive filtering scheme and
algorithms based on an alternating optimization strategy with
shrinkage. The proposed scheme employs a two-stage structure that
consists of an alternating optimization of a diagonally-structured
matrix that accelerates the convergence and an adaptive filter with
a shrinkage function that attracts the coefficients with small
magnitudes to zero. The diagonally-structure matrix aims to perform
what the oracle algorithm does and helps to accelerate the
convergence of the scheme and improve its steady-state performance.
We devise sparsity-aware alternating optimization least-mean square
(SA-ALT-LMS) algorithms for the proposed scheme and derive
analytical formulas to predict their mean-square error (MSE) upon
convergence. Simulations for a system identification application
show that the proposed scheme and algorithms outperform in
convergence and tracking the state-of-the-art sparsity-aware
algorithms.

\section{Problem Statement and The Oracle Algorithm}

In this section, we state the sparse system identification problem
and describe the optimal strategy known as as the oracle algorithm,
which knows the positions of the non-zero coefficients of the sparse
system.

\subsection{Sparse System Identification Problem}


In the sparse system identification problem of interest, 
the system observes a complex-valued signal represented by an $M
\times 1$ vector ${\boldsymbol x}[i]$ at time instant $i$, performs
filtering and obtains the output $d[i] = {\boldsymbol
w}^H_o{\boldsymbol x}[i]$, where ${\boldsymbol w}_o$ is an
$M$-length finite-impulse-response (FIR) filter that represents the
actual system. For system identification, an adaptive filter with
$M$ coefficients ${\boldsymbol w}[i]$ is employed in such a way that
it observes ${\boldsymbol x}[i]$ and produces an estimate
$\hat{d}[i] = {\boldsymbol w}^H[i]{\boldsymbol x}[i]$. The system
identification scheme then compares the output of the actual system
$d[i]$ and the adaptive filter $\hat{d}[i]$, resulting in an error
signal $e[i] = d[i] + n[i] - \hat{d}[i]$, where $n[i]$ is the
measurement noise. In this context, the goal of an adaptive
algorithm is to identify the system by minimizing the MSE defined by
\begin{equation}
{\boldsymbol w}_o = \arg \min_{\boldsymbol w} E[|d[i] + n[i] -
{\boldsymbol w}^H[i] {\boldsymbol x}[i]|^2]
\end{equation}
A key problem in electronic measurement systems which are modeled by
sparse adaptive filters, where the number of non-zero coefficients
$K << M$, is that most adaptive algorithms do not exploit their
sparse structure to obtain performance benefits and/or a
computational complexity reduction. If an adaptive algorithm can
identify and exploit the non-zero coefficients of the system to be
identified, then it can obtain performance improvements and a
reduction in the computational complexity.

\subsection{The Oracle Algorithm}

The optimal algorithm for processing sparse signals and systems is
known as the oracle algorithm. It can identify the positions of the
non-zero coefficients and fully exploit the sparsity of the system
under consideration. In the context of sparse system identification
and other linear filtering problems, we can state the oracle
algorithm as
\begin{equation}
\{ {\boldsymbol P}_{\rm or}, {\boldsymbol w}_{\rm or} \} = \arg
\min_{{\boldsymbol P}, {\boldsymbol w}} E[|d[i] + n[i] -
{\boldsymbol w}^H{\boldsymbol P}{\boldsymbol x}[i] |^2]
\end{equation}
where ${\boldsymbol P}_{\rm or}$  is an $M \times M$ diagonal
matrix with the actual $K$ positions of the non-zero coefficients.
It turns out that the oracle algorithm requires an exhaustive search
over all the possible $K$ positions over $M$ possibilities, which is
an $NP$-hard problem with extremely high complexity if $M$ is large.
Moreover, the oracle algorithm also requires the computation of the
optimal filter, which is a continuous optimization problem. For
these reasons, it is fundamental to devise low-complexity algorithms
that can cost-effectively process sparse signals.

\section{Proposed Alternating Optimization with Shrinkage Scheme}

\begin{figure}[!htb]
\begin{center}
\def\epsfsize#1#2{0.7\columnwidth}
\epsfbox{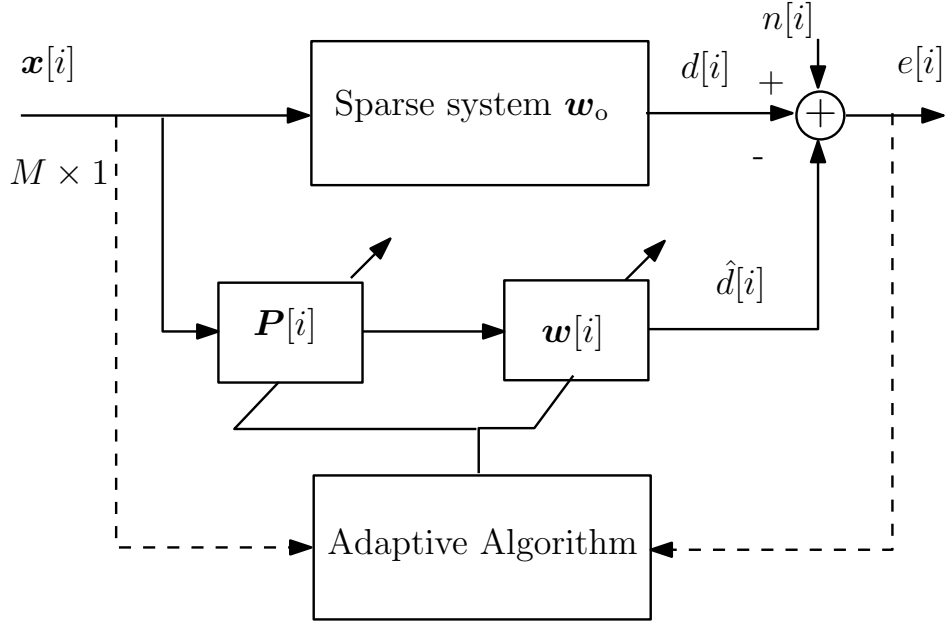} \vspace{-0.5em}\caption{ Proposed adaptive
filtering scheme.}\label{fig:alt}
\end{center}
\end{figure}

In this section, we present an adaptive filtering scheme that
employs an alternating optimization strategy with shrinkage that
exploits the sparsity in the identification of linear systems.
Unlike existing methods, the proposed technique introduces two
adaptive filters that are optimized in an alternating fashion, as
illustrated in Fig. \ref{fig:alt}. The first adaptive filter
${\boldsymbol p}[i]$ with $M$ coefficients is applied as a diagonal
matrix ${\boldsymbol P}[i] = {\rm diag}({\boldsymbol p}[i])$ to
${\boldsymbol x}[i]$ and performs the role of the oracle algorithm,
which was defined as ${\boldsymbol P}_{\rm or}$ in the previous
section. The second adaptive filter ${\boldsymbol w}[i]$ with $M$
coefficients is responsible for the system identification. Both
${\boldsymbol p}[i]$ and ${\boldsymbol w}[i]$ employ $l_1$-norm
shrinkage techniques to attract to zero the coefficients that have
small magnitudes.  The output of the proposed adaptive filtering
scheme is given by
\begin{equation}
\begin{split}
\hat{d}[i] & = {\boldsymbol w}^H[i] \underbrace{{\boldsymbol P}[i]}_{{\rm diag}({\boldsymbol p}[i])}
{\boldsymbol x}[i] = {\boldsymbol p}^T[i] \underbrace{{\boldsymbol W}^*[i]}_{{\rm diag}({\boldsymbol w}^*[i])}
{\boldsymbol x}[i] \\
& = {\boldsymbol x}^T[i] {\boldsymbol P}[i] {\boldsymbol w}^*[i] = {\boldsymbol x}^T[i] {\boldsymbol W}^*[i]{\boldsymbol p}[i]
\end{split}
\end{equation}

\subsection{Adaptive Algorithms}

In order to devise adaptive algorithms for this scheme, we need to
cast an optimization problem with a cost function that depends on
${\boldsymbol p}[i]$, ${\boldsymbol w}[i]$ and a shrinkage function
$f(\cdot)$,  where $f({\boldsymbol a})$ represents this function
applied to a generic parameter vector ${\boldsymbol a}$ with $M$
coefficients. Let us consider the following cost function
\begin{equation}
\begin{split}
C({\boldsymbol p}[i], {\boldsymbol w}[i]) & = E[|d[i] -
\hat{d}[i]|^2] + \lambda f({\boldsymbol p}[i]) )+ \tau
f({\boldsymbol w}[i]), \label{cost}
\end{split}
\end{equation}
where $\lambda$ and $\tau$ are the regularization terms. In order to
derive an adaptive algorithm to minimize the cost function in
(\ref{cost}) and perform system identification, we employ an
alternating optimization strategy. We compute the instantaneous
gradient of (\ref{cost}) with respect to ${\boldsymbol p}[i]$ and
${\boldsymbol w}[i]$ and devise LMS-type algorithms:
\vspace{-0.25em}
\begin{equation}
\begin{split}
\hspace{-0.5em}{\boldsymbol p}[i+1] & = {\boldsymbol p}[i] - \eta
\lambda \frac{\partial C({\boldsymbol p}[i], {\boldsymbol
w}[i])}{\partial {\boldsymbol
p}^*[i] } \\
& ={\boldsymbol p}[i] + \eta e({\boldsymbol w}[i],{\boldsymbol
p}[i]) {\boldsymbol W}[i] {\boldsymbol x}^*[i] - \underbrace{\eta
\lambda}_{\alpha} \frac{\partial f({\boldsymbol p}[i])}{\partial
{\boldsymbol p}^*[i] },\label{prec}
\end{split}
\end{equation}
\vspace{-0.75em}
\begin{equation}
\begin{split}
\hspace{-0.5em}{\boldsymbol w}[i+1] & = {\boldsymbol w}[i] - \mu
\frac{\partial C({\boldsymbol p}[i], {\boldsymbol w}[i])}{\partial
{\boldsymbol
w}^*[i] } \\
& ={\boldsymbol w}[i] + \mu e({\boldsymbol w}[i],{\boldsymbol
p}[i])^* {\boldsymbol P}[i] {\boldsymbol x}[i] - \underbrace{\mu
\tau}_{\gamma} \frac{\partial f({\boldsymbol w}[i])}{\partial
{\boldsymbol w}^*[i] } ,\label{wrec}
\end{split}
\end{equation}
where $e({\boldsymbol w}[i],{\boldsymbol p}[i]) = d[i] -
{\boldsymbol w}^H[i] {\boldsymbol P}[i] {\boldsymbol x}[i]$ is the
error signal and $\mu$ and $\eta$ are the step sizes of the LMS
recursions, which are used in an alternating way. In Table I,
different shrinkage functions are shown with their partial
derivatives and other features. {A key requirement of the proposed
scheme is the initialization which results in the adjustment of
${\boldsymbol p}[i]$ to shrink the coefficients corresponding to
zero elements of the system and ${\boldsymbol w}[i]$ to estimate the
non-zero coefficients. Specifically, ${\boldsymbol p}[i]$ is
initialized as an all-one vector (${\boldsymbol p}[0]= {\boldsymbol
1}$ or ${\boldsymbol P}[0] = {\boldsymbol I}$) and ${\boldsymbol
w}[i]$ is initialized as an all-zero vector (${\boldsymbol w}[0]=
{\boldsymbol 0}$ ). When ${\boldsymbol p}[i]$ is fixed, the scheme
is equivalent to a standard shrinkage algorithm. The two-step
approach outperforms the single-step method since ${\bf P}[i]$
strives to perform the role of the Oracle algorithm (${\bf P}_{\rm
or}$) by decreasing the values of its entries in the positions of
the zero coefficients. This helps the recursion that adapts ${\bf
w}[i]$ to perform the estimation of the non-zero coefficients. This
process is then alternated over the iterations, resulting in better
performance. When ${\bf P}_{\rm or}$ is employed, ${\bf w}[i]$ has
the information about the actual positions of the zero coefficients.
}

%
%

\begin{table*}
\centering \vspace{-0.5em} \caption{Shrinkage functions}
\vspace{-1.25em} \footnotesize
\begin{tabular}{llll}
\hline  Function & Partial Derivative  & ${\boldsymbol
L}_{\boldsymbol a}$ & Cost of Shrinkage ($C_s$)\\
\hline
 $ {\small f({\boldsymbol a})=||{\boldsymbol a}||_{1}}$ & ${{\small \frac{\partial f({\boldsymbol a}[i])}{\partial
{\boldsymbol a}^*[i] }={\rm sgn}({\boldsymbol a})}= {\rm sgn}(\Re[
{\boldsymbol a}]) + j {\rm sgn}(\Im[{\boldsymbol a}])}$ & ${\small
\approx {\rm sgn}[ {\boldsymbol a}_{\rm opt}]{\rm sgn}[ {\boldsymbol
a}_{\rm opt}^H]} $ & ${\footnotesize 2M {\rm ad} + 4M{\rm mult} +
2M{\rm div} }$
 \\ \\
$ {\small f({\boldsymbol a})= \sum_{m=1}^M \log (1+ |a_m|/\epsilon)
}$ & ${\small \frac{\partial f({\boldsymbol a}[i])}{\partial
{\boldsymbol a}^*[i] }=\frac{{\rm sgn}(\Re[{\boldsymbol a}]) + j
{\rm sgn}(\Im [ {\boldsymbol a}])}{1+ \epsilon ||{\boldsymbol
a}||_{1}}}$ & ${\small \approx \frac{{\rm sgn}[ {\boldsymbol a}_{\rm
opt}]}{1+\epsilon|{\boldsymbol a}_{\rm opt} |}\frac{{\rm
sgn}[ {\boldsymbol a}_{\rm opt}^H]}{1+\epsilon|{\boldsymbol a}_{\rm opt} |}}$  & ${\small 4M {\rm ad} + 7M{\rm mult} + 3M{\rm div}}$\\ \\
${\small f({\boldsymbol a}) = ||a||_0 }$ & ${\small \frac{\partial
f({ a}_m[i])}{\partial { a}^*_m[i] }=   \begin{cases}
    \beta \big( {\rm sgn}(\Re[a_m]) +   & \text{if } |a_m| \leq 1/\beta \\
     j{\rm sgn}(\Im[a_m]) \big)- \beta^2 a_m & \\
    0       & {\rm elsewhere}
   \end{cases}}$ & ${\small \approx \beta^2 {\rm sgn}[ {\boldsymbol a}_{\rm opt}]{\rm sgn}[ {\boldsymbol a}_{\rm opt}^H]
    }$ & ${\small 3M {\rm ad} + 6M{\rm mult} + 2M{\rm div}}$ \\
   ${\small \approx \sum_{m=1}^{M} (1 - e^{-\beta |a_m|})}$ & & ${\small - \beta^3 {\rm sgn}[{\boldsymbol a}_{\rm opt}]{\boldsymbol a}_{\rm opt}^H }$   &
\\
    & & ${\small - \beta^3 {\boldsymbol a}_{\rm opt}{\rm sgn}[{\boldsymbol a}_{\rm opt}^H] }$   &
\\
    & & ${\small + \beta^4 {\boldsymbol a}_{\rm opt}{\boldsymbol a}_{\rm opt}^H }$   &

\\
\hline
\end{tabular}
\label{table1}
\end{table*}

\subsection{Computational Complexity}

We detail the computational complexity in terms of arithmetic
operations of the proposed and some existing algorithms.
Specifically, we consider the conventional LMS algorithm,
sparsity-aware LMS (SA-LMS) algorithms, and the proposed SA-ALT-LMS
algorithm. The details are shown in Table \ref{table2}.

\begin{table}[h]
\centering \small \vspace{-0.75em} \caption{Computational Complexity
of Algorithms} \vspace{-1.0em}
\begin{tabular}{ll}
\hline
Algorithm & Computational Complexity \\
\hline
 LMS & $2M{\rm ad} + 2M{\rm mult}$ \\
SA-LMS & $2M{\rm ad} + 2M{\rm mult} + 2C_s$\\
SA-ALT-LMS & $5M{\rm ad} + 7M{\rm mult} + 2C_s$ \\
\hline
\end{tabular}
\label{table2}
\end{table}
\vspace{-1.0em}

\section{Mean-Square Error Analysis}

In this section, we develop an MSE analysis of the proposed
SA-ALT-LMS algorithm and devise analytical expressions to describe
the transient and steady-state performances. By defining
${\boldsymbol w}_{\rm o}$ as the optimal filter and ${\boldsymbol
p}_{\rm o}$ as the oracle vector (${\boldsymbol P}_{\rm o} = {\rm
diag}({\boldsymbol p}_{\rm o}))$ with the non-zero coefficients, we
can write
\begin{equation}
{\boldsymbol e}_{w} = {\boldsymbol w}[i] - { \boldsymbol w}_{\rm o} ~~ {\rm and} ~~
{\boldsymbol e}_{p} = {\boldsymbol p}[i] - {\boldsymbol p}_{\rm o}.
\end{equation}
The error signal can then be rewritten as
\begin{equation}
\begin{split}
e({\boldsymbol w}[i],{\boldsymbol p}[i]) & = e_{\rm o} -
{\boldsymbol x}^T[i]( {\rm diag}({\boldsymbol e}_{\rm p}^T[i])
{\boldsymbol e}_{w}^*[i] \\ & \quad + {\rm diag}( {\boldsymbol
e}_{p}^T[i]) {\boldsymbol w}_{\rm o}^* + {\rm diag} ({\boldsymbol
p}_{\rm o}^T) {\boldsymbol e}_{w}^*[i]),
\end{split}
\end{equation}
where $e_{\rm o}=e({\boldsymbol w}_{\rm o}, {\boldsymbol p}_o) =
d[i] - {\boldsymbol x}^T[i] {\rm diag} ({\boldsymbol p}_{\rm o})^T
{\boldsymbol w}_{\rm o}^*$ is the error signal of the optimal sparse
filter. The MSE is written as
\begin{equation}
\begin{split}
{\rm MSE} & = E[ |e({\boldsymbol w}[i],{\boldsymbol p}[i])|^2] \\
& = E[|e_{\rm o} - {\boldsymbol x}^T[i]( {\rm diag}({\boldsymbol
e}_{\rm p}^T[i]) {\boldsymbol e}_{w}^*[i] \\ & \quad + {\rm diag}(
{\boldsymbol e}_{p}^T[i]) {\boldsymbol w}_{\rm o}^* + {\rm diag}
({\boldsymbol p}_{\rm o}^T) {\boldsymbol e}_{w}^*[i])|^2]
\end{split}
\end{equation}
Using the independence assumption between ${\boldsymbol e}_{p}[i]$,
${\boldsymbol e}_{w}[i]$ and ${\boldsymbol x}[i]$, we have:
\begin{equation}
\begin{split}
{\rm MSE} & = J_{\rm min} + E[{\boldsymbol x}^H[i] {\rm
diag}({\boldsymbol e}_{p}^H) {\boldsymbol e}_{w}[i]{\boldsymbol
e}_{w}^H[i] {\rm diag} ({\boldsymbol e}_{p}[i])
{\boldsymbol x}[i]] \\
& \quad +  E[{\boldsymbol x}^H[i] {\rm diag}({\boldsymbol e}_{p}^H)
{\boldsymbol w}_{\rm o}[i]{\boldsymbol w}_{\rm o}^H[i] {\rm diag}
({\boldsymbol e}^{p}[i])
{\boldsymbol x}[i]] \\
& \quad +  E[{\boldsymbol x}^H[i] {\rm diag}({\boldsymbol p}_{\rm
o}^H) {\boldsymbol e}_{w}[i]{\boldsymbol e}_{w}^H[i] {\rm diag}
({\boldsymbol p}_{\rm o}[i]) {\boldsymbol x}[i]],
\end{split}
\end{equation}
where $J_{\rm min} = E[|e({\boldsymbol w}_o,{\boldsymbol p}_o)|^2]$.
The expectation of the scalar values that are functions of triple
vector products can be rewritten \cite{haykin} and the MSE expressed
by
\begin{equation}
\begin{split}
{\rm MSE} & = J_{\rm min} + {\rm tr}[{\boldsymbol R}_x ({\boldsymbol
K}_w \odot {\boldsymbol K}_p) ] \\ & \quad + {\rm tr}[{\boldsymbol
R}_x ({\boldsymbol R}_{w_{\rm o}} \odot {\boldsymbol K}_{p})] + {\rm
tr}[{\boldsymbol R}_x ( {\boldsymbol R}_{\rm or} \odot {\boldsymbol
K}_{w})], \label{mse_trace}
\end{split}
\end{equation}
where $\odot$ is the Hadamard product, ${\boldsymbol
R}_x=E[{\boldsymbol x}[i]{\boldsymbol x}^H[i]]$, ${\boldsymbol K}_w
= E[{\boldsymbol e}_w[i]{\boldsymbol e}_{w}^H[i]]$, ${\boldsymbol
K}_{p} = E[{\boldsymbol e}_{p}[i] {\boldsymbol e}_{p}^H[i]]$,
${\boldsymbol R}_{w_{\rm o}} = E[{\boldsymbol w}_{\rm o}
{\boldsymbol w}_{\rm o}^H]$, and ${\boldsymbol R}_{\rm or} =
E[{\boldsymbol p}_{\rm o}{\boldsymbol p}_{\rm o}^H]$. Using
(\ref{prec}) and (\ref{wrec}) into ${\boldsymbol K}_{p}$ and
${\boldsymbol K}_{w}$, we obtain
\begin{equation}
\begin{split}
{\boldsymbol K}_{w}[i+1] & = ({\boldsymbol I} - \mu{\boldsymbol
R}_{px}){\boldsymbol K}_{w}[i]({\boldsymbol I} - \mu{\boldsymbol
R}_{px}) \\ & \quad+ \mu^2 {\boldsymbol R}_{px}J_{\rm
MSE}^{(i)}({\boldsymbol w}_{\rm o})  + \gamma^2 {\boldsymbol L}_w,
\end{split}
\end{equation}
\begin{equation}
\begin{split}
{\boldsymbol K}_{p}[i+1] & = ({\boldsymbol I} - \eta {\boldsymbol
R}_{wx}){\boldsymbol K}_{p}[i]({\boldsymbol I} - \eta {\boldsymbol
R}_{wx}) \\ & \quad + \eta^2 {\boldsymbol R}_{wx}J_{\rm
MSE}^{(i)}({\boldsymbol p}_{\rm o})  + \alpha^2 {\boldsymbol L}_p,
\end{split}
\end{equation}
where $J_{\rm MSE}^{(i)}({\boldsymbol w}_o) \triangleq
E[|e({\boldsymbol w}_o, {\boldsymbol p}[i])|^2]$ and $J_{\rm
MSE}^{(i)}({\boldsymbol p}_o) \triangleq E[|e({\boldsymbol w}[i],
{\boldsymbol p}_o)|^2]$ appear in (12) and (13). The other
quantities are ${\boldsymbol R}_{wx} = E[{\boldsymbol
W}[i]{\boldsymbol x}[i] {\boldsymbol x}^H[i] {\boldsymbol W}^H[i]]$,
${\boldsymbol L}_w = E[f'[{\boldsymbol w}[i]] f'^H[{\boldsymbol
w}[i]]$, ${\boldsymbol R}_{px} = E[{\boldsymbol P}[i]{\boldsymbol
x}[i] {\boldsymbol x}^H[i] {\boldsymbol P}^H[i]]$, ${\boldsymbol
L}_p = E[f'[{\boldsymbol p}[i]] f'^H[{\boldsymbol p}[i]]$ and
$f'(\cdot)$ is the partial derivative with respect to the variable
of the argument. In Table I, we use the variable ${\boldsymbol a}$
that plays the role of ${\boldsymbol p}[i]$ or ${\boldsymbol w}[i]$.
We obtained approximations for ${\boldsymbol L}_a =
E[f'[{\boldsymbol a}[i]] f'^H[{\boldsymbol a}[i]]$, where
${\boldsymbol a}$ is a generic function, to compute the matrices
${\boldsymbol L}_p$ and ${\boldsymbol L}_w$ for a given shrinkage
function as shown in the $3$rd column of Table I.

We can express ${\boldsymbol R}_{wx}$ and ${\boldsymbol R}_{px}$ as
${\boldsymbol R}_{wx} ={\boldsymbol R}_x \odot {\boldsymbol
R}_w[i]]$ and ${\boldsymbol R}_{px} ={\boldsymbol R}_x \odot
{\boldsymbol R}_p[i]$, where ${\boldsymbol R}_{w} = E[{\boldsymbol
w}[i]{\boldsymbol w}^H[i]]$ and ${\boldsymbol R}_{p} =
E[{\boldsymbol p}[i]{\boldsymbol p}^H[i]]$. To simplify the
analysis, we assume that the samples of the signal ${\boldsymbol
x}[i]$ are uncorrelated, i.e., ${\boldsymbol R}_x = \sigma_x^2
{\boldsymbol I}$ with $\sigma_x^2$ being the variance. Using the
diagonal matrices ${\boldsymbol R}_x = {\boldsymbol \Lambda}_x
=\sigma_x^2 {\boldsymbol I}$, ${\boldsymbol R}_{px} = {\boldsymbol
\Lambda}_{px}[i] = \sigma_x^2 {\boldsymbol I} \odot {\boldsymbol
R}_p[i]$ and ${\boldsymbol R}_{wx} = \Lambda_{wx}[i] = \sigma_x^2
{\boldsymbol I} \odot {\boldsymbol R}_w[i]$, we can write
\begin{equation}
\begin{split}
{\boldsymbol K}_{w} [i+1] & = ({\boldsymbol I} - \mu {\boldsymbol
\Lambda}_{px}[i]){\boldsymbol K}_{w} [i] ({\boldsymbol I} - \mu
{\boldsymbol \Lambda}_{px}[i]) \\ & \quad + \mu^2 J_{\rm
MSE}^{(i)}({\boldsymbol w}_{o}) {\boldsymbol \Lambda}_{px}[i] +
\gamma^2 {\boldsymbol L}_w[i]
\end{split}
\end{equation}
\begin{equation}
\begin{split}
{\boldsymbol K}_{p} [i+1] & = ({\boldsymbol I} - \eta {\boldsymbol
\Lambda}_{wx}[i]){\boldsymbol K}_{p} [i] ({\boldsymbol I} - \eta
{\boldsymbol \Lambda}_{wx}[i]) \\ & \quad + \eta^2 J_{\rm
MSE}^{(i)}({\boldsymbol p}_{o}) {\boldsymbol \Lambda}_{wx}[i] +
\alpha^2 {\boldsymbol L}_{p}[i]
\end{split}
\end{equation}
Due to the structure of the above equations, the approximations and
the quantities involved, we can decouple them into
\begin{equation}
\begin{split}
{K}_{w}^n [i+1] & = (1 - \mu { \lambda}_{px}^n[i]){K}_{w}^n [i] (1 -
\mu {\lambda}_{px}^n[i])
\\ & \quad + \mu^2 J_{\rm MSE}^{(i)}({\boldsymbol w}_{o})
{\lambda}_{px}^n[i] + \gamma^2 { L}_{w}^{n}[i]
\end{split}
\end{equation}
\begin{equation}
\begin{split}
{K}_{p}^n [i+1] & = (1 - \eta \lambda_{wx}^n[i]){
K}_{p}^n [i] (1 - \eta \lambda_{wx}^n[i]) \\
& \quad + \eta^2 J_{\rm MSE}^{(i)}({\boldsymbol p}_{o})
\lambda_{wx}[i] + \alpha^2 L_p^{n}[i]
\end{split}
\end{equation}
where $K_w^n[i]$ and $K_p^n[i]$ are the $n$th elements of the main
diagonals of ${\boldsymbol K}_w[i]$ and ${\boldsymbol K}_p[i]$,
respectively. By taking $\lim_{i \rightarrow \infty} K_{w}^n[i+1]$
and $\lim_{i \rightarrow \infty} K_{p}^n[i+1]$, we obtain
\begin{equation}
K_{w}^n = \frac{J({\boldsymbol w}_o)}{(2/\mu - \lambda_{px}^n)} +
\frac{\gamma^2 { L}_w^{n}}{\mu^2 \lambda_{px}^n(2/\mu -
\lambda_{px}^n)}
\end{equation}
\begin{equation}
K_{p}^n = \frac{J({\boldsymbol p}_o)}{(2/\eta - \lambda_{wx}^n)} +
\frac{\alpha^2 { L}_p^n}{\eta^2 \lambda_{wx}^n(2/\eta -
\lambda_{wx}^n)},
\end{equation}
where $J({\boldsymbol w}_o) = \lim_{i \rightarrow \infty} J_{\rm
MSE}^{(i)}({\boldsymbol w}_o)$ and $J({\boldsymbol p}_o) = \lim_{i
\rightarrow \infty} J_{\rm MSE}^{(i)}({\boldsymbol p}_o)$. For
stability, we must have $|1- \mu \lambda_{x}^n|<1$ and $|1-\eta
d_{x}^n | < 1$, which results in
\begin{equation}
0< \mu < 2/\max_{n} [\lambda_{px}^n] ~~{\rm and}~~0< \eta <
2/\max_{n} [\lambda_{wx}^n],
\end{equation}
where $\lambda_{px}^n = \lim_{i \rightarrow \infty} \sigma_x^2
E[|p^n[i]|^2]$, $\lambda_{wx}^n = \lim_{i \rightarrow \infty}
\sigma_x^2 E[|w^n[i]|^2]$, with $p^n[i]$ and $w^n[i]$ being the
$n$th elements of ${\boldsymbol p}[i]$ and ${\boldsymbol w}[i]$,
respectively. The MSE is then given by
\begin{equation}
\begin{split}
{\rm MSE} & = J_{\rm min} +  \sigma_x^2 \sum_{n=1}^M K_p^n   M_{w}^n  \\
& \quad+ \sigma_x^2 \sum_{n=1}^M   p_o^n |w_o^n|^2 K_{p}^n
 + \sigma_x^2 \sum_{n=1}^M p_o^n  K_{w}^n,
\label{mse_anal}
\end{split}
\end{equation}
where $w_o^{n}$ and $p_{o}^n$ are the elements of ${\boldsymbol
w}_{\rm o}$ and ${\boldsymbol p}_{\rm o}$, respectively. This MSE
analysis is valid for uncorrelated input data, whereas a model for
correlated input data remains an open problem which is highly
involved due to the triple products in (\ref{mse_trace}). However,
the SA-ALT-LMS algorithms work very well for both correlated and
uncorrelated input data.

\section{Simulations}

In this section, we assess the performance of the existing LMS,
SA-LMS, and the proposed SA-ALT-LMS algorithms with different
shrinkage functions. The shrinkage functions considered are the ones
shown in Table II, which give rise to the SA-LMS with the $l_1$-norm
\cite{chen}, {the SA-LMS with the log-sum penalty
\cite{chen,eksioglu,candes} and the $l_0$-norm \cite{gu,eksioglu2}.}
We consider system identification examples with both time-invariant
and time-varying parameters in which there is a sparse system with a
significant number of zeros to be identified. The input signal
${\boldsymbol x}[i]$ and the noise ${\boldsymbol n}[i]$ are drawn
from independent and identically distributed complex Gaussian random
variables with zero mean and variances $\sigma_x^2$ and
$\sigma_n^2$, respectively, resulting in a signal-to-noise ratio
(SNR) given by ${\rm SNR} =\sigma_x^2/\sigma_n^2$. {The filters are
initialized as ${\boldsymbol p}[0]= {\boldsymbol 1}$ and
${\boldsymbol w}[0]= {\boldsymbol 0}$.} In the first experiment,
there are $N=16$ coefficients in a time-invariant system, only $K=2$
coefficients are non-zero when the algorithms start and the input
signal is applied to a first-order auto-regressive filter which
results in correlated samples obtained by $x_c[i] = 0.8x_c[i-1] +
x[i]$ that are normalized. After $1000$ iterations, the sparse
system is suddenly changed to a system with $N=16$ coefficients but
in which $K=4$ coefficients are non-zero. The positions of the
non-zero coefficients are chosen randomly for each independent
simulation trial. The curves are averaged over $200$ independent
trials and the parameters are optimized for each example. {We
consider the log-sum penalty \cite{chen,eksioglu,candes} and the
$l_0$-norm \cite{gu,eksioglu2} because they have shown the best
performances.}

\begin{figure}[!htb]
\begin{center}
\def\epsfsize#1#2{0.875\columnwidth}
\epsfbox{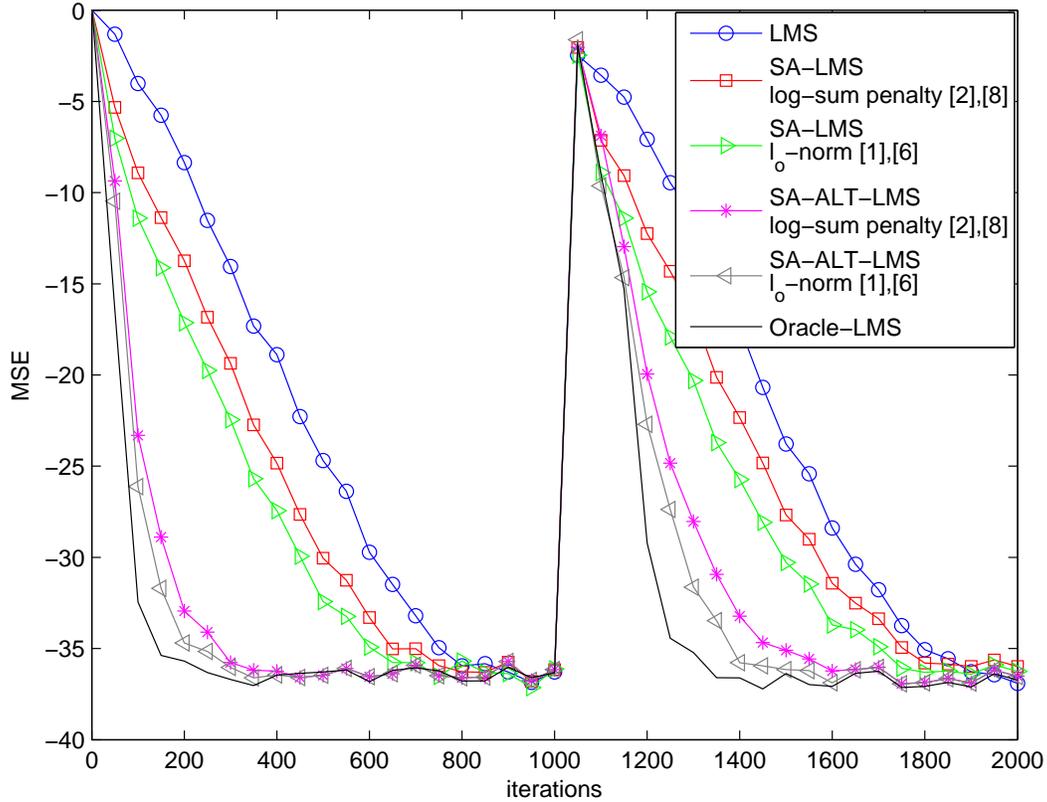} \vspace{-1.5em}\caption{MSE performance against
number of iterations for correlated input data. Parameters: ${\rm
SNR}=40 dB$, $\sigma_x^2=1$, $\mu = 0.015$, $\eta = 0.012$, $\tau =
0.02$, $\lambda=0.02$, $\epsilon = 10$, and $\beta = 10$.
}\label{fig2}
\end{center}
\end{figure}

The results of the first experiment are shown in Fig. \ref{fig2},
where the existing LMS and SA-LMS algorithms are compared with the
proposed SA-ALT-LMS algorithm. The curves show that that MSE
performance of the proposed SA-ALT-LMS algorithms is significantly
superior to the existing LMS and SA-LMS algorithms for the
identification of sparse system. The SA-ALT-LMS algorithms can
approach the performance of the Oracle-LMS algorithm, which has full
knowledge about the positions of the non-zero coefficients. A
performance close to the Oracle-LMS algorithm was verified for
various situations of interest including different values of SNR,
degrees of sparsity ($K$) and for both small and large sparse
systems ($10 \leq N \leq 200$).

\begin{figure}[!htb]
\begin{center}
\def\epsfsize#1#2{0.875\columnwidth}
\epsfbox{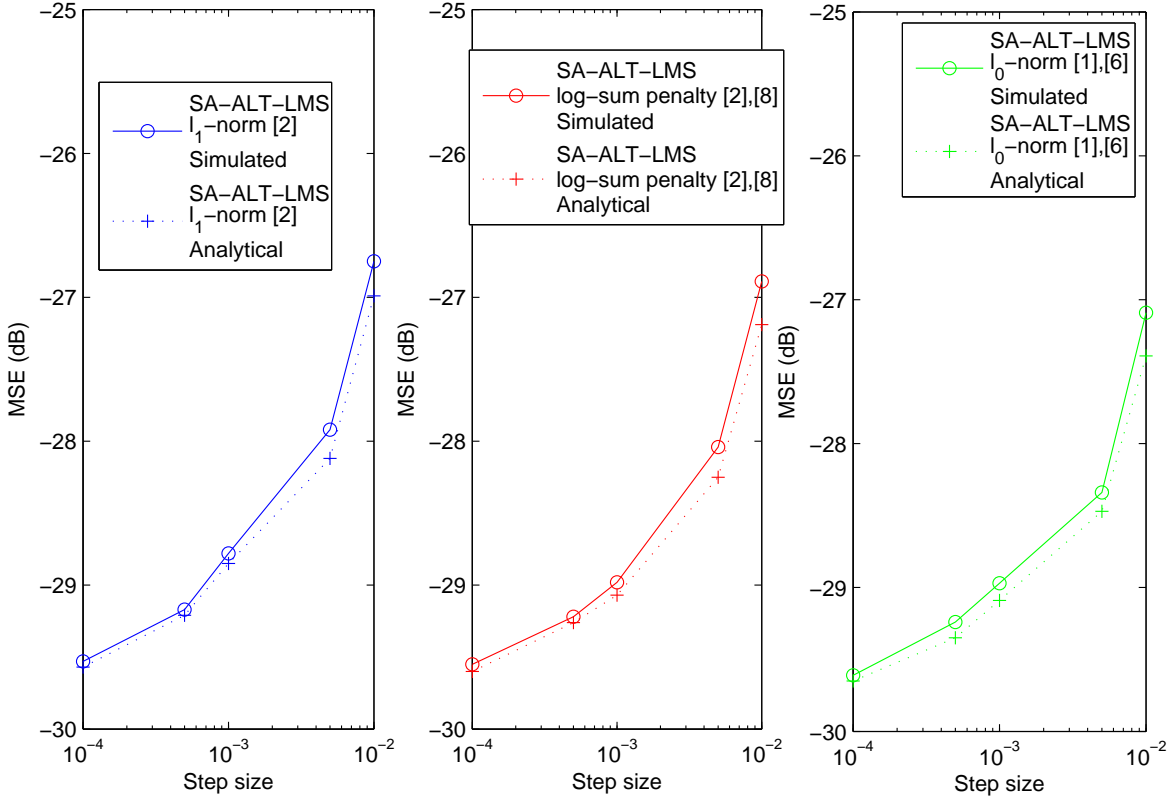} \vspace{-1.35em}\caption{MSE performance against
step size for $\mu=\eta$. Parameters: ${\rm SNR}=30 dB$,
$\sigma_x^2=1$, $\tau = 0.02$, $\lambda=0.02$, $\epsilon = 10$, and
$\beta = 10$. }\label{fig3}
\end{center}
\end{figure}

In a second experiment, we have assessed the validity of the MSE
analysis and the formulas obtained to predict the MSE as indicated
in (\ref{mse_anal}) and in Table II for uncorrelated input data. In
the evaluation of (18) and (19), we made the following
approximations $J({\boldsymbol w}_o) \approx J({\boldsymbol p}_o)
\approx J_{\rm min}$, $\lambda_{px}^n \approx \sigma_x^2 p_o^n$ and
$\lambda_{wx}^n \approx \sigma_x^2 w_o^n$. We have considered a
scenario where the input signal and the observed noise are white
Gaussian random sequences with variance of $1$ and $10^{-3}$,
respectively, i.e., ${\rm SNR} = 30$ dB. There are $N=32$
coefficients in a time-invariant system that are randomly generated
and only $K=4$ coefficients are non-zero. The positions of the
non-zero coefficients are again chosen randomly for each independent
simulation trial. The curves are averaged over $200$ independent
trials and the algorithms operate for $1000$ iterations in order to
ensure their convergence. We have compared the simulated curves
obtained with the SA-ALT-LMS strategy using the $l_1$-norm
\cite{chen}, {the SA-LMS with the log-sum penalty
\cite{chen,eksioglu,candes} and the $l_0$-norm \cite{gu,eksioglu2}.}
The results in Fig. \ref{fig3} indicate that there is a close match
between the simulated and the analytical curves for the shrinkage
functions employed, suggesting that the formulas obtained and the
simplifications made are valid and resulted in accurate methods to
predict the MSE performance of the proposed SA-ALT-LMS algorithms.

\section{Conclusion}

We have proposed a novel sparsity-aware adaptive filtering
scheme and algorithms based on an alternating optimization strategy
that is general and can operate with different shrinkage functions.
We have devised alternating optimization LMS
algorithms, termed as SA-ALT-LMS for the proposed scheme and
developed an MSE analysis, which resulted in analytical formulas
that can predict the performance of the SA-ALT-LMS algorithms.
Simulations for a system identification application
show that the proposed scheme and SA-ALT-LMS algorithms outperform
existing sparsity-aware algorithms.

\end{document}